**Computable Phenotypes of Patient Acuity in the Intensive Care Unit**

Yuanfang Ren, PhD[1,2], Jeremy Balch, MD[3], Kenneth L. Abbott, MD[3], Tyler J. Loftus, MD[1,3], Benjamin Shickel, PhD[1,2], Parisa Rashidi, PhD[1,4], Azra Bihorac, MD MS[1,2], Tezcan Ozrazgat-Baslanti, PhD[1,2]

[1] Intelligent Clinical Care Center (IC[3]), University of Florida, Gainesville, FL, USA

[2] Department of Medicine, College of Medicine, University of Florida, Gainesville, FL, USA

[3] Department of Surgery, College of Medicine, University of Florida, Gainesville, FL, USA

[4] J. Crayton Pruitt Family Department of Biomedical Engineering, University of Florida, Gainesville, FL, USA

Correspondence to: Azra Bihorac MD MS, Intelligent Clinical Care Center (IC[3]), Department of Medicine, Division of Nephrology, Hypertension, and Renal Transplantation, PO Box 100224, Gainesville, FL 32610-0224. Telephone: (352) 294-8580; Fax: (352) 392-5465; Email: abihorac@ufl.edu

Reprints will not be available from the author(s).

Short title: Computable phenotypes for patient acuity status

Conflict of Interest Disclosures: None reported.

Funding/Support: Conflicts of Interest and Sources of Funding: P.R., A.B., T.O.B., and Y.R were supported by R01 EB029699 from the National Institute of Biomedical Imaging and Bioengineering (NIH/NIBIB), This work was supported in part by the National Institutes of Health and National Center for Advancing Translational Sciences Clinical and Translational Sciences Award UL1 grant TR000064. The content is solely the responsibility of the authors and does not necessarily represent the official views of the National Institutes of Health. The funders had no role in study design, data collection and analysis, decision to publish, or manuscript preparation. A.B. and T.O.B. had full access to all study data and take responsibility for data integrity and data analysis.

**Abstract**

**Background:** Continuous monitoring and patient acuity assessments are key aspects of Intensive Care Unit (ICU) practice, but both are limited by time constraints imposed on healthcare providers. Moreover, anticipating clinical trajectories remains imprecise. The objectives of this study are to (1) develop an electronic phenotype of acuity using automated variable retrieval within the electronic health records and (2) describe transitions between acuity states that illustrate the clinical trajectories of ICU patients.

**Methods:** We gathered two single-center, longitudinal electronic health record datasets for 51,372 adult ICU patients admitted to the University of Florida Health (UFH) Gainesville (GNV) and Jacksonville (JAX). Records included demographic information, comorbidities, vital signs, results of laboratory studies, medications, and diagnoses and procedure codes for all index admissions, as well as any admissions within 12 months before or after index admissions. We developed algorithms to quantify acuity status at four-hour intervals for each ICU admission and identify acuity phenotypes using continuous acuity status and k-means clustering approach.

**Results:** 51,073 admissions for 38,749 patients in the UFH GNV dataset and 22,219 admissions for 12,623 patients in the UFH JAX dataset had at least one ICU stay lasting more than four hours. Compared to UFH JAX cohort with an ICU admission, patients admitted to UFH GNV ICU had worse clinical outcomes. There were three phenotypes: persistently stable, persistently unstable, and transitioning from unstable to stable. Transitions occurred almost exclusively between 12 and 56 hours after ICU admission. For stable patients, approximately 0.7%-1.7% would transition to unstable, 0.02%-0.1% would expire, 1.2%-3.4% would be discharged, and the remaining 96%-97% would remain stable in the ICU every four hours. For unstable patients, approximately 6%-10% would transition to stable, 0.4%-0.5% would expire, and the remaining 89%-93% would remain unstable in the ICU in the next four hours.

**Conclusions:** We developed phenotyping algorithms for patient acuity status every four hours while admitted to the ICU. This approach may be useful in developing prognostic and clinical decision-support tools to aid patients, caregivers, and providers in shared decision-making processes regarding escalation of care and patient values.

## INTRODUCTION

Each year, 5.7 million patients are admitted to intensive care units (ICUs) in the United States, with ICU mortality ranging from 10-29% and costs exceeding $82 billion, representing more than 4.1% of national health expenditures.[1 2] Continuous monitoring and dynamic patient acuity assessments are key aspects of ICU care, but both are limited by the time constraints imposed on healthcare providers. ICU patient acuity assessments rely heavily on physicians' clinical judgment and vigilance. Yet, ICU physicians spend only 9.4% of their clinical time in direct patient contact.[3] Similarly, most ICU nurses spend approximately 30-50% of their time on direct patientcare, suggesting that patients may not be directly observed by a care provider for a substantial portion of their ICU stay.[4] In addition, manual observations suffer from subjectivity, poor recall, and limited number of administrations per day, leading to missed opportunities for time-sensitive interventions.[5-9] Thus, there is a critical, unmet need for automated, data-driven ICU patient assessments.

Using electronic health record data for two retrospective, longitudinal cohorts of ICU patients, we developed automated, computable phenotypes for acuity status and described transitions among acuity states that illustrate the clinical trajectories of ICU patients.

## METHODS

### Data source and participants

Using the University of Florida Health (UFH) Integrated Data Repository as an honest broker, we gathered two single-center, longitudinal electronic health record (EHR) datasets for all adult patients admitted to UFH Gainesville (GNV) and Jacksonville (JAX). The UFH GNV dataset included 383,193 hospital admissions for 121,800 patients between June 1, 2014 and August 22, 2019, and the UFH JAX dataset included 283,310 hospital admissions for 60,047 patients between June 1, 2014 and May 1, 2021. We focused on admissions with at least one ICU stay lasting more than four hours (eFigure 1). The UFH GNV dataset included 51,073 ICU admissions for 38,749 patients meeting this criterion, and the UFH JAX dataset included 22,219 ICU admissions for 12,623 patients meeting this criterion.

Each dataset included demographic information, comorbidities, vital signs, laboratory values, medications, and diagnoses and procedure codes for all index admissions as well as admissions within 12 months before and after index admissions. All EHR were de-identified, except that dates of service were maintained. This study was approved by the University of Florida Institutional Review Board and Privacy Office (IRB#201600223, IRB#201600262 and IRB#201901123) as an exempt study with a waiver of informed consent.

### Definition of acuity status

At the end of each four-hour interval during an ICU stay, patient acuity status was classified as stable, unstable, expired, or discharged alive, as illustrated in Figure 1. Patients were considered unstable if they required at least one of the following life-support therapies: vasopressors, mechanical ventilation, continuous renal replacement therapy, or massive blood transfusion (defined as transfusion of at least 10 units of red blood cells within the prior 24 hours). All other patients who remained in the ICU at the end of the four-hour interval were classified as stable.

### Data elements and rules used for identification of instability

Data elements used for identification of acuity status are provided in eTable 1. Computable phenotyping algorithms used a rule-based approach, as detailed below.

**Identification of ICU stay:** We identified the dates and times during which a patient was assigned an ICU bed. Patient movements within the hospital (e.g., to radiology and operating room suites) often generate separate electronic health record encounters despite remaining assigned to an ICU bed, therefore, ICU stays within 24 hours of one another were merged.

**Identification of vasopressor use:** Use of vasopressors was identified by the presence of at least one of the vasopressor RxNorm[10] codes listed in eTable 2 for intravenous epinephrine, vasopressin, phenylephrine, norepinephrine, or dopamine administered during the four-hour interval of interest.

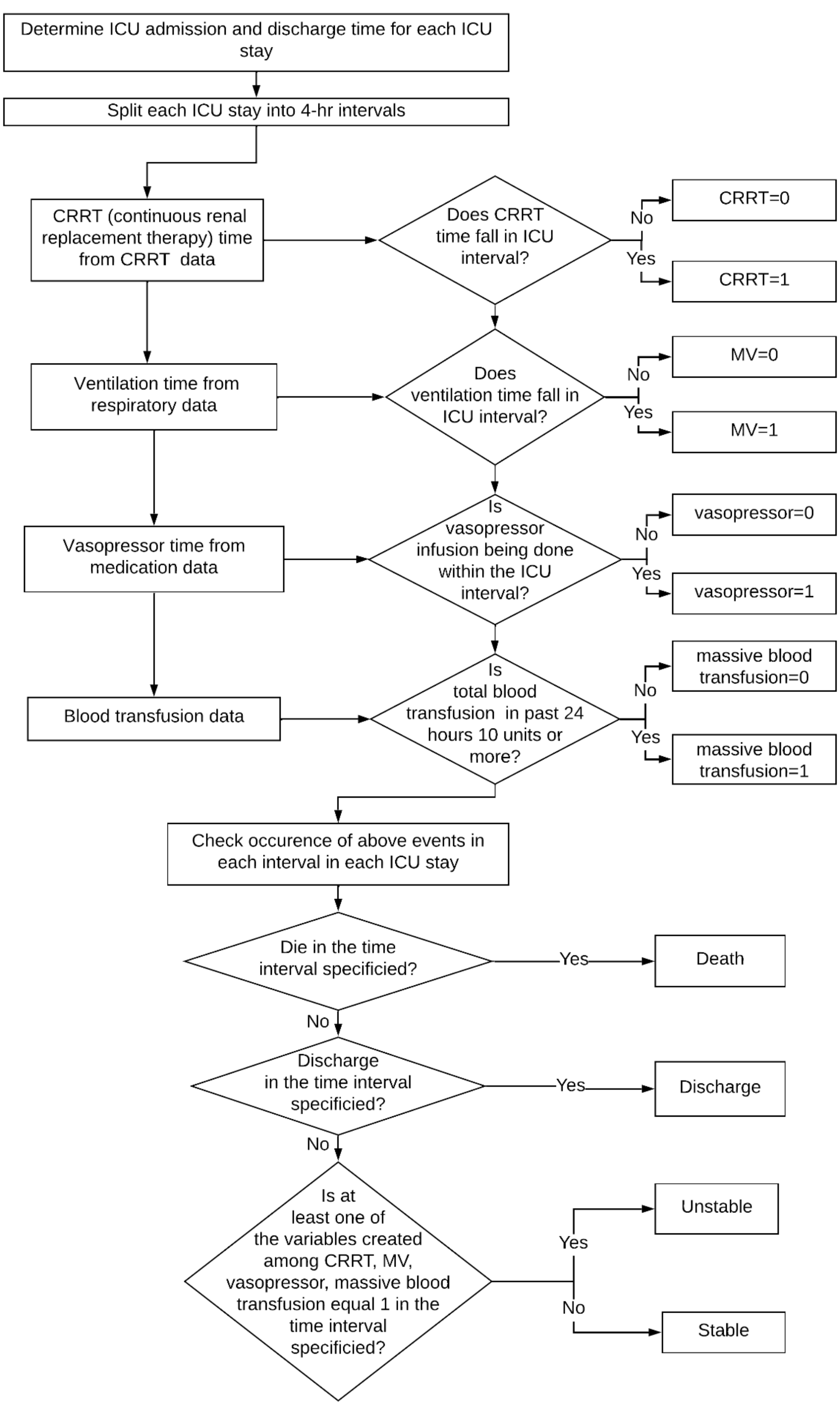

**Figure 1. Flowchart for phenotyping ICU patient acuity status.**

**Identification of continuous renal replacement therapy:** We identified continuous renal replacement therapy (CRRT) using EHR date and time stamps and orders specifying CRRT treatment type. We identified timeframes during which the treatment type was continuous veno-venous hemofiltration (CVVH), continuous veno-venous hemodialysis (CVVHD), or continuous veno-venous hemodiafiltration (CVVHDF).

**Identification of massive blood transfusion:** For each four-hour interval, we determined whether the total amount of red blood cell transfusion in the prior 24 hours met or exceeded 10 units, as illustrated in eFigure 2. These analyses included transfusions for which either the start or end date and time stamps occurred within the 24-hour interval of interest. If the number of units for a transfusion order was not missing, but either the start or end date and time stamp was missing, the missing date and time stamp was imputed based on the median duration of transfusion in the cohort. If both start and end date and time stamps were missing, the start date and time was imputed as the median interval between order placement and initiation of the blood transfusion in the cohort, which was 1.6 hours.

**Identification of mechanical ventilation:** We identified use of mechanical ventilation using EHR data representing respiratory devices, ventilation modes, and measured values for respiratory vitals that include oxygen flow rate, tidal volume, and positive end-expiratory pressure, as illustrated in eFigure 3 and eTable 3. Patients were assumed to be on mechanical ventilation if at least one of the mechanical ventilation parameters was present or the respiratory device was identified as an invasive ventilator. For patients undergoing surgery, which often requires general anesthesia and mechanical ventilation unrelated to respiratory failure, we used the last available respiratory data prior to the start of surgery.

## Statistic methods

To understand how patient acuity status evolves in the ICU, we first presented the clinical course of patients via alluvia plots where we stratified the patients by considering their movements between the acuity states within the first three days in ICU. We then applied *k*-means clustering to derive acuity transition patterns using time series acuity status (stable or unstable) within three days of ICU admission. For admissions lasting less than three days, we characterized missing acuity status as stable for patients who were discharged alive and as unstable for patients who expired within three days of ICU admission. Thus, for each ICU admission, we had 18 (3 * 24 / 4) acuity data points to be analyzed by k-means clustering algorithms. We developed our model on UFH GNV dataset and assessed reproducibility by re-deriving clusters using *k*-means clustering algorithm on the UFH JAX dataset. We compared the frequency distributions and clinical outcomes, including length of stay in the hospital, length of stay in the ICU, and in hospital mortality across phenotypes.

To ensure the acuity status was not simply a recapitulation of traditional illness severity measures, we tested whether the acuity status was explained by Sequential Organ Failure Assessment (SOFA).[11] SOFA score was calculated every four hours using EHR data within the last 24 hours. We grouped SOFA score into two classes: SOFA score less than or equal to 6 was termed "mild SOFA"; SOFA score greater than 6 was termed "severe SOFA". We presented the SOFA score distribution and transition using alluvia plots. We also derived the SOFA score transition patterns by applying *k*-means clustering approach to time series SOFA scores (mild or severe SOFA) within the first three days of ICU admission.

For comparison, $\chi 2$ test for categorical variables and the Kruskal-Wallis test for continuous variables were performed. We adjusted p values for the family-wise error rate due to multiple comparisons using the Bonferroni correction. Statistical analyses were performed with Python version 3.7 and R version 4.3.1.

## RESULTS

## Patients

Clinical characteristics and outcomes for patients were assessed for both cohorts (Table 1). Among 51,073 patient encounters with at least one ICU admission in the UFH GNV cohort, the mean (SD) age was 59 (17) years; 22,929 (45%) encounters were female; 38,136 (75%) encounters were White and 9,707 (19%) were African American; 1,974 (4%) encounters were Hispanic and 48,343 (95%) were non-Hispanic. The UFH JAX cohort contain 22,219 patient encounters with ICU admission and the demographic characteristics was significantly different from UFH GNV cohort (mean (SD) age, 59 (14) years; 10,414 (47%) female 10,996 (49%) White and 10,308 (46%) African American; 584 (3%) Hispanic and 21,542 (97%) non-Hispanic). Compared with those not admitted to the ICU, patients in both cohorts requiring ICU admission had longer median hospital stay (UFH GNV: 7 days vs. 1 day; UFH JAX: 4 days vs. 1 day) and higher in-hospital mortality (UFH GNV: 10% vs. 0.4%; UFH JAX: 3% vs. 0.7%). For patients who were admitted to the ICU and subsequently expired, more deaths occurred in the ICU than on general hospital wards (UFH GNV: 7% vs. 1%; UFH JAX: 1.4% vs. 0.7%). Compared to the UFH JAX ICU cohort, patients admitted to UFH GNV ICU had worse clinical outcomes including longer median hospital stay (7 days vs. 4 day), longer median ICU stay (4 days vs. 3 days), and higher in-hospital mortality (10% vs. 3%).

### Distribution of acuity status

Distributions of acuity status for two cohorts within the initial 15 days of ICU admission were illustrated in Figure 2A and eFigure 4A. In the UFH GNV cohort, the number of patients still admitted to the ICU decreased rapidly from over 50,000 to 10,000 in seven days, marking the 75$^{th}$ percentile for ICU days (Figure 2A). Acuity status was generally stable or unstable. Of the patients remaining in the ICU there were comparatively few discharges and deaths. The proportion of patients with stable acuity status increased within first 48 hours after ICU admission but then gradually decreased for approximately two weeks. The maximum percentages of patients who were discharged alive or expired during each four-hour period were 1.8% and 0.4%, respectively.

**Table 1. Summary of patient characteristics.**

| | UF Health GNV Admissions (06/01/2014-08/22/2019) | UF Health JAX Admissions (06/01/2014-05/01/2021) | *p*-values |
|---|---|---|---|
| Number of patients | 121,800 | 60,047 | |
| Number of encounters | 383,193 | 283,310 | |
| Number of patients with at least one ICU stay | 38,749 | 12,623 | |
| Number of encounters with at least one ICU stay | 51,073 | 22,219 | |
| Number of ICU stays | 54,178 | 22,530 | |
| **Preadmission clinical characteristics** of ICU patients | | | |
| Age, years, mean (SD) | 59 (17) | 59 (14) | **0.003** |
| Female sex, n (%) | 22,929 (45) | 10,414 (47) | **<0.001** |
| Race, n (%) | | | |
| White | 38,136 (75) | 10,996 (49) | **0** |
| African American | 9,707 (19) | 10,308 (46) | **0** |
| Other | 2,547 (5) | 828 (4) | **<0.001** |
| Missing | 683 (1) | 87 (0.4) | **<0.001** |
| Ethnicity, n (%) | | | |
| Hispanic | 1,974 (4) | 584 (3) | **<0.001** |
| Non-Hispanic | 48,343 (95) | 21,542 (97) | **<0.001** |
| Missing | 756 (1) | 93 (0.4) | **<0.001** |
| **Hospital admissions, n** | **383,193** | **283,310** | |
| Hospital LOS, days, median (IQR) | 2 (1, 4) | 1 (1, 3) | **0** |
| In-hospital mortality, n (%) | 6,269 (2) | 2,436 (1) | **<0.001** |
| **Admissions not involving an ICU stay, n (%)** | **332,120 (87)** | **261,091 (92)** | |
| Hospital LOS, days, median (IQR) | 1 (1, 3) | 1 (1, 3) | **0** |
| In-hospitality mortality, n (%) | 1,362 (0.4) | 1,872 (0.7) | **<0.001** |
| **ICU admissions, n (%)** | **51,073 (13)** | **22,219 (8)** | |
| Hospital LOS, days, median (IQR) | 7 (4, 13) | 4 (2, 8) | **0** |
| ICU LOS, days, median (IQR) | 4 (2, 7) | 3 (2, 5) | **<0.001** |
| In-hospitality mortality, n (%) | 4,907 (10) | 564 (3) | **<0.001** |
| Death within 7-days of ICU admission, n (%) | 2,991 (6) | 282 (1) | **<0.001** |
| Percentage of intervals with stable status from ICU admission to hospital discharge, %, median (IQR) | 100 (87, 100) | 100 (100, 100) | **0** |
| **ICU admissions with death in ICU, n (%)** | **3,777 (7)** | **300 (1)** | |
| Hospital LOS, days, median (IQR) | 6 (3, 14) | 6 (2,15) | 0.90 |
| ICU LOS, days, median (IQR) | 4 (2, 10) | 5 (2, 11) | 0.82 |
| Death within 7-days of ICU admission, n (%) | 2,535 (67) | 194 (65) | 0.42 |
| Percentage of intervals with stable status from ICU admission to hospital discharge, %, median (IQR) | 17 (2, 54) | 56 (3, 97) | **<0.001** |
| **ICU admissions with death on general ward, n (%)** | **389 (1)** | **148 (1)** | |
| Hospital LOS, days, median (IQR) | 12 (7, 20) | 16 (9, 28) | **0.002** |
| ICU LOS, days, median (IQR) | 5 (3, 9) | 4 (2, 7) | **<0.001** |
| Death within 7-days of ICU admission, n (%) | 161 (41) | 40 (27) | **0.003** |
| Percentage of intervals with stable status from ICU admission to hospital discharge, %, median (IQR) | 84 (48, 100) | 63 (26, 91) | **<0.001** |

Abbreviations: ICU: intensive care unit; LOS: length of stay; SD: standard deviation; IQR: interquartile range.

Compared to the UFH GNV cohort, members of the UFH JAX cohort were more likely to present in stable status (eFigure 4A). The proportion of patients with stable acuity remained above 80%, significantly higher than the UFH GNV cohort. The maximum percentages of patients who were discharged alive or expired during each four-hour period were 7.5% and 0.3%, respectively.

**Acuity status transitions**

Acuity status transition probabilities were illustrated in Figure 2B and eFigure 4B. In the UFH GNV cohort, for stable patients, approximately 1.7% would transition to unstable, 0.1% would expire, 1.2% would be discharged, and 97% would remain stable in the ICU every four hours. For unstable patients, every four hours, approximately 6% would transition to stable, 0.5% would expire, and the remaining 93% would remain unstable in the ICU. Compared to the UFH GNV cohort, members of the UFH JAX cohort were more likely to transition from unstable to stable status (10% vs. 6%), less likely to transition to unstable status from stable status (0.7% vs. 1.7%) and more likely to be discharged (3.4% vs. 1.2%).

Dynamic acuity status transition within the first three days of ICU admission was demonstrated in Figure 2C-D and eFigure 4C-D. Among patients admitted with stable status, a large proportion either remained stable or were discharged, with negligible numbers of patients progressing to unstable status. Patients admitted in unstable status where most likely to remain unstable, with gradually increasing numbers of transition to stability and constant rates of transition to death.

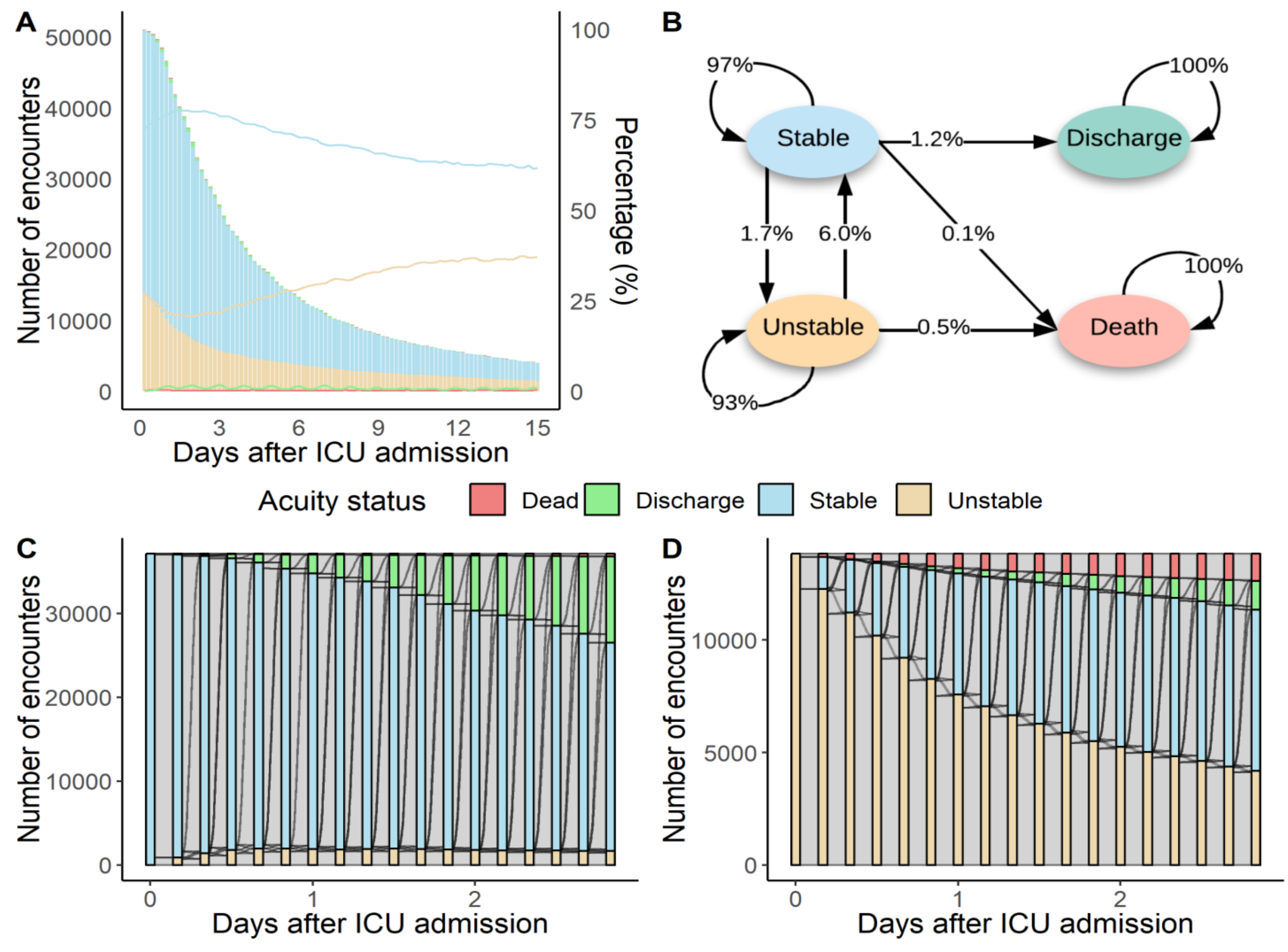

**Figure 2. Distribution and transition of acuity status for the UFH GNV cohort.** (A) Distribution of acuity status within 15 days of ICU admission. The left y-axis shows the total number of encounters for all acuity states at each time point, which is represented as a bar graph. The right y-axis shows the percentage of each acuity state at each time point, which is represented as a line graph. (B) Acuity state transition probability in every four-hour interval. (C) Acuity status transition for patients with stable status on ICU admission. (D) Acuity status transition for patients with unstable status on ICU admission.

Clustering analyses of the UFH GNV cohort (Figure 3A) and the UFH JAX cohort (eFigure 5A) suggested three acuity status phenotypes during the first 72 hours of ICU admission. Cluster 1 was the largest group (UFH GNV 37,180, 73% and UFH JAX 21,095, 95%, Table 2 and eTable 4), characterized by persistent stability, the lowest in-hospital mortality (UFH GNV 4% and UFH GNV 1%), and the shortest median length of ICU (UFH GNV 3 days and UFH JAX 3 days) and hospital stay (UFH GNV 6 days and UFH JAX 4 days). Cluster 2 initially presented with instability with quick transition to stability, occurring almost exclusively in the first 12 and 56 hours after ICU admission. This cluster had the smallest proportion of patients (UFH GNV 6,219, 12% and UFH JAX 436, 2%), intermediate in-hospital mortality (UFH GNV 5% and UFH GNV 3%) and intermediate lengths of stay in the ICU (UFH GNV 4 days and UFH JAX 4 days) and hospital (UFH GNV 9 days and UFH JAX 7 days). Cluster 3 was characterized by persistent instability with an intermediate proportion of patients (UFH GNV 7,674, 15% and UFH JAX 688, 3%), the highest in-hospital mortality (UFH GNV 41% and UFH GNV 35%), and longest median length of stay in the ICU (UFH GNV 8 days and UFH JAX 6 days) and hospital (UFH GNV 13 days and UFH JAX 11 days).

**Comparison to SOFA score**

To test whether the acuity status was explained by SOFA score, k-means clustering was also conducted using the time series SOFA score that was calculated every 4 hours, within the first 72 hours (Figure 3B and eFigure 5B). The SOFA score exhibited similar transition patterns with acuity status. The transition time point for Cluster 2 was however several hours later than that for acuity status Cluster 2. The cluster frequency distribution and clinical outcome of SOFA

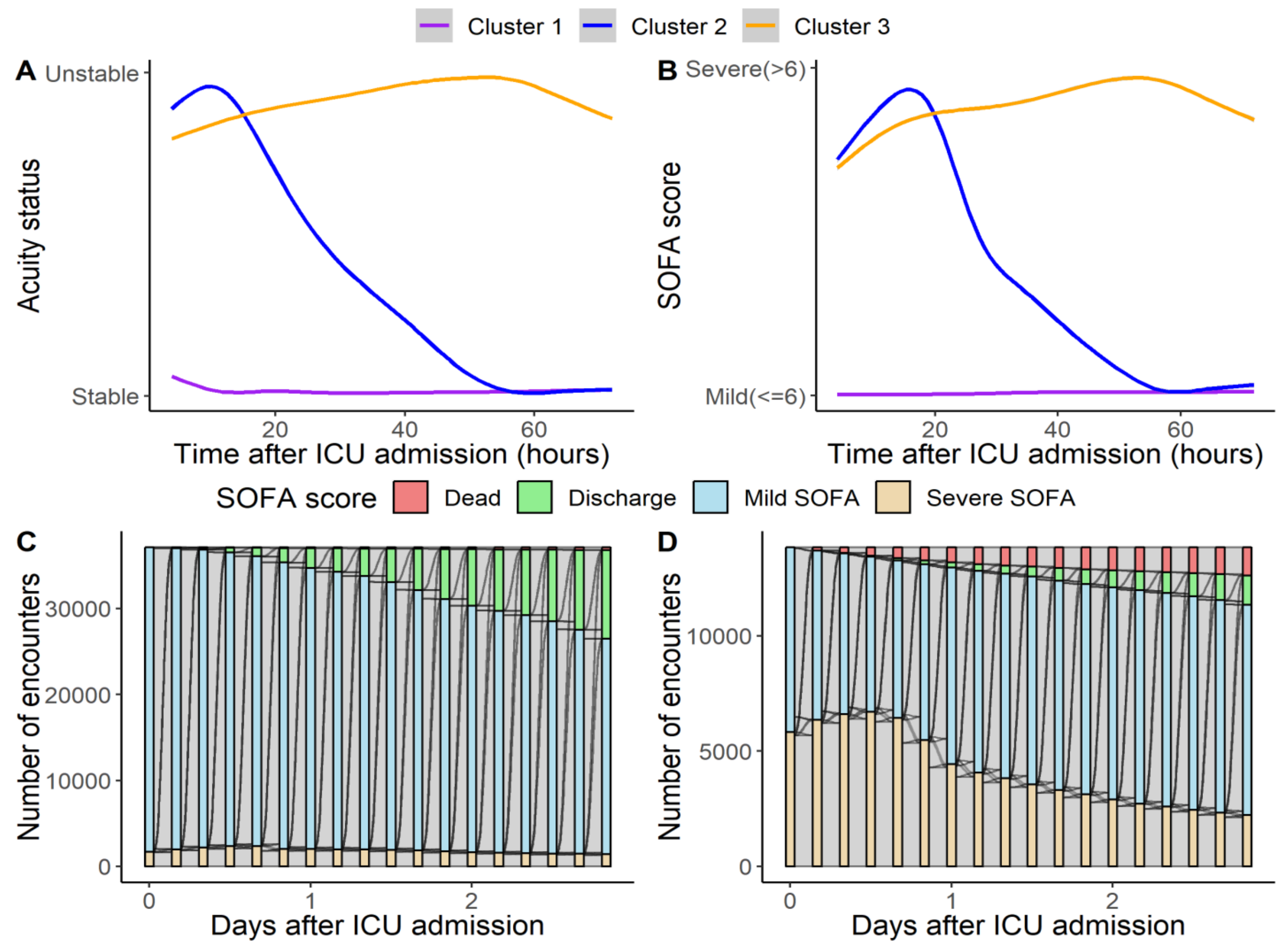

**Figure 3. Phenotypes of acuity status transition and comparison to Sequential Organ Failure Assessment (SOFA) score for the UFH GNV cohort.** (A) Phenotypes of acuity status transition within the first three days of ICU admission. (B) Phenotypes of SOFA score transition within the first three days of ICU admission. (C) SOFA score transition for patients with stable status on ICU admission. (D) SOFA score transition for patients with unstable status on ICU admission.

**Table 2. Clinical outcomes across phenotypes of acuity status and Sequential Organ Failure Assessment (SOFA) score on UFH GNV cohort.**

| **Acuity status phenotypes** | **Cluster 1** | **Cluster 2** | **Cluster 3** |
|---|---|---|---|
| Number of encounters, n (%) | 37,180 (73) | 6,219 (12) | 7,674 (15) |
| Length of hospital stay, days, median (IQR) | 6 (4, 11) | 9 (6, 14)[a] | 13 (6, 22)[a,b] |
| Length of ICU stay, days, median (IQR) | 3 (2, 5) | 4 (3, 7)[a] | 8 (4, 16)[a,b] |
| Hospital mortality, n (%) | 1,468 (4) | 304 (5)[a] | 3,135 (41)[a,b] |
| Death within 7 days of ICU admission, n (%) | 638 (2) | 134 (2) | 2, 219 (29)[a,b] |
| **SOFA score phenotypes** | **Cluster 1** | **Cluster 2** | **Cluster 3** |
| Number of encounters, n (%) | 30,408 (60) | 16,455 (32) | 4,210 (8) |
| Length of hospital stay, days, median (IQR) | 6 (4, 10) | 9 (6, 16)[a] | 12 (4, 22)[a] |
| Length of ICU stay, days, median (IQR) | 3 (2, 5) | 5 (3, 9)[a] | 7 (3, 15)[a,b] |
| Hospital mortality, n (%) | 699 (2) | 2064 (13)[a] | 2066 (49)[a,b] |
| Death within 7 days of ICU admission, n (%) | 287 (0.9) | 1124 (7)[a] | 1580 (38)[a,b] |

Abbreviations: SOFA: sequential organ failure assessment; ICU: intensive care unit; IQR: interquartile range.
All p-values were adjusted for multiple comparisons using Bonferroni method.
[a] $p < 0.05$ compared to Cluster 1. .
[b] $p < 0.05$ compared to Cluster 2.

score phenotypes were different from acuity status phenotypes (Table 2). SOFA score Cluster 2 contained the second highest number of patients and had higher hospital mortality in the UFH GNV cohort compared to Cluster 2 of acuity status (UFH GNV 13% vs. 5% and UFH JAX 3% vs. 3%). Both acuity status phenotypes (Cluster 2 and Cluster 3) contained a substantial proportion of patients with mild and severe SOFA score within first 24 hours of ICU admission (eFigure 6 and eFigure 7) and were not dominated by one SOFA score group. Clinical outcomes among patients with mild SOFA score across the three acuity status phenotypes were presented in eTable 5. Patients with mild SOFA score in acuity status Cluster 2 and 3 had longer length of hospital stay and ICU stay. Patients in Cluster 3 had significantly higher mortality compared to other clusters.

Dynamic SOFA score transition within the first three days of ICU admission was demonstrated in Figure 3C-D and eFigure 5C-D. Over 90% of patients admitted with stable status presented with mild SOFA score. Unstable patients, however, presented with a near even mix of mild SOFA score and severe SOFA scores. And the percentage of patients with high SOFA score initially increased within the first 16 hours and gradually decreased. The percentage of patients with severe SOFA score was consistent lower than that of patients with unstable status (Figure 2D, Figure 3D, eFigure 4D and eFigure 5D).

The SOFA score transition pattern is significantly different from acuity state transition pattern, demonstrating that acuity status is not explained by SOFA score.

**DISCUSSION**

Our results mirror intuition regarding the clinical trajectory of ICU patients. The first 24 to 48 hours after ICU admission saw an increase in the raw number and percentage of ICU patients who were stable. After 48 hours, rates of discharge from the ICU, including death, remained constant though most individual patients continued to transition from instability to stability. Understandably, as these stable patients were discharged from the ICU, an increasing proportion of the remaining ICU population was unstable. This trend continued for approximately two weeks after ICU admission, consistent with clinical definitions of chronic critical illness.[12,13] Also intuitively, the probability of death during the following four-hour period was approximately four-fold higher among unstable vs. stable patients.

For both stable and unstable patients, there was a 7% probability that the acuity state would be different four hours later. Through clustering analyses using dynamic acuity status, we identified three ICU patient phenotypes: patients who were persistently stable, patients who were persistently unstable, and patients who were initially unstable but became stable within 56 hours. Patients who were persistently unstable had significantly higher mortality and longer length of ICU and hospital stay than other phenotypes, suggesting that identifying these phenotypes early has the potential to allow for patient risk stratification.

Such observations involving ICU patient phenotypes may inform prognostic discussions among patients, caregivers, and providers. For unstable critical illness, patients and their caregivers may wish to embark on a course of aggressive, life-sustaining treatments if there is a high probability of recovery and transition to stability at the time of ICU admission. Some critically ill patients may have previously expressed a desire to forego prolonged life-sustaining treatments. In this case, there may be utility in providing the patient and their caregivers with accurate, data-driven predictions that the probability of early recovery is low. While SOFA is also predictive of mortality, acuity status transition may provide an additional data point for clinical decision making and plan of care discussions. This information could augment the decision-making process and alleviate the stress associated with the decision to forgo aggressive, resource-intense therapy.

Our methods for phenotyping can be automated. Manual and repetitive patient assessments contribute to personnel shortages and burnout. Critical care teams are under significant work pressure,[14] and almost a third of ICU nursing teams suffer from burnout.[15] High nursing workload may contribute to life-threatening adverse events in the ICU[16-20] and highlights the need for automation of routine tasks.[21,22] Autonomous assessments can enhance critical care workflow efficiency by facilitating routine nursing assessments in the ICU and allow nurses to spend time on more critical tasks. In addition, assessments that are associated with prognosis and clinical trajectory have the potential to augment prognostication and decisions regarding escalation of care and resource use.

This study has several limitations. First, our use of data from only two institutions limits the generalizability of our findings to other institutions. Second, we made binary distinctions between stable and unstable patients to facilitate identification of transitions in care. However, we recognize that true patient acuity exists on a continuum. Finally, whether such identification of acuity states improves prognostication and clinical decision-making remains unknown and should be the subject of future research.

## CONCLUSION

We developed phenotyping algorithms that determined patient acuity status every four hours during ICU admission. This task can be automated, which has the advantage of avoiding additional patient assessments by ICU health care workers, who already face worsening work force shortages and job-related stress and burnout. Automated acuity phenotyping has the potential to leverage high-resolution physiological signals and digital EHR data to develop prognostic and clinical decision-support tools that aid patients, caregivers, and providers in shared decision-making processes regarding escalation of care and patient values.

**REFERENCES**

1. Wallace DJ, Angus DC, Seymour CW, Barnato AE, Kahn JM. Critical care bed growth in the United States. A comparison of regional and national trends. *Am J Respir Crit Care Med* 2015; 191(4): 410-6.
2. Elias KM, Moromizato T, Gibbons FK, Christopher KB. Derivation and validation of the acute organ failure score to predict outcome in critically ill patients: a cohort study. *Crit Care Med* 2015; 43(4): 856-64.
3. Mamykina L, Vawdrey DK, Hripcsak G. How Do Residents Spend Their Shift Time? A Time and Motion Study With a Particular Focus on the Use of Computers. *Acad Med* 2016; 91(6): 827-32.
4. Wong DH, Gallegos Y, Weinger MB, Clack S, Slagle J, Anderson CT. Changes in intensive care unit nurse task activity after installation of a third-generation intensive care unit information system. *Crit Care Med* 2003; 31(10): 2488-94.
5. Brown H, Terrence J, Vasquez P, Bates DW, Zimlichman E. Continuous monitoring in an inpatient medical-surgical unit: a controlled clinical trial. *Am J Med* 2014; 127(3): 226-32.
6. Kipnis E, Ramsingh D, Bhargava M, et al. Monitoring in the intensive care. *Crit Care Res Pract* 2012; 2012: 473507.
7. Wollschlager CM, Conrad AR, Khan FA. Common complications in critically ill patients. *Dis Mon* 1988; 34(5): 221-93.
8. Rubins HB, Moskowitz MA. Complications of care in a medical intensive care unit. *J Gen Intern Med* 1990; 5(2): 104-9.
9. Desai SV, Law TJ, Needham DM. Long-term complications of critical care. *Crit Care Med* 2011; 39(2): 371-9.
10. Nelson SJ, Zeng K, Kilbourne J, Powell T, Moore R. Normalized names for clinical drugs: RxNorm at 6 years. *J Am Med Inform Assoc* 2011; 18(4): 441-8.
11. Vincent JL, Moreno R, Takala J, et al. The SOFA (Sepsis-related Organ Failure Assessment) score to describe organ dysfunction/failure. On behalf of the Working Group on Sepsis-Related Problems of the European Society of Intensive Care Medicine. *Intensive Care Med* 1996; 22(7): 707-10.
12. Iwashyna TJ, Hodgson CL, Pilcher D, et al. Timing of onset and burden of persistent critical illness in Australia and New Zealand: a retrospective, population-based, observational study. *Lancet Respir Med* 2016; 4(7): 566-73.
13. Kahn JM, Le T, Angus DC, et al. The epidemiology of chronic critical illness in the United States*. *Crit Care Med* 2015; 43(2): 282-7.
14. Coomber S, Todd C, Park G, Baxter P, Firth-Cozens J, Shore S. Stress in UK intensive care unit doctors. *Br J Anaesth* 2002; 89(6): 873-81.
15. Verdon M, Merlani P, Perneger T, Ricou B. Burnout in a surgical ICU team. *Intensive Care Med* 2008; 34(1): 152-6.
16. Bucknall TK. Medical error and decision making: Learning from the past and present in intensive care. *Aust Crit Care* 2010; 23(3): 150-6.
17. Ksouri H, Balanant PY, Tadie JM, et al. Impact of morbidity and mortality conferences on analysis of mortality and critical events in intensive care practice. *Am J Crit Care* 2010; 19(2): 135-45; quiz 46.

18. Pagnamenta A, Rabito G, Arosio A, et al. Adverse event reporting in adult intensive care units and the impact of a multifaceted intervention on drug-related adverse events. *Ann Intensive Care* 2012; 2(1): 47.
19. Valentin A, Capuzzo M, Guidet B, et al. Patient safety in intensive care: results from the multinational Sentinel Events Evaluation (SEE) study. *Intensive Care Med* 2006; 32(10): 1591-8.
20. Valentin A, Schiffinger M, Steyrer J, Huber C, Strunk G. Safety climate reduces medication and dislodgement errors in routine intensive care practice. *Intensive Care Med* 2013; 39(3): 391-8.
21. Jastremski CA. Retention of critical care nurses: Important to the future of critical care. *Crit Care Med* 2006; 34(7): 2015.
22. Duke EM. The Critical Care Workforce: A Study of the Supply and Demand for Critical Care Physicians. *Health Resouree and Service Administration US Department of Health and Human Services W ashington DC US A 2008* 2006.

**Supplementary Online Content**

**Ren Y, Balch J, Abbott K, Loftus T, Shickel B, Rashidi P, Bihorac A, Ozrazgat-Baslanti T.**
**Computable Phenotypes of Patient Acuity in the Intensive Care Unit**

This supplementary material has been provided by the authors to give readers additional information about their work.

## Supplemental Material Table of Contents

**eFigure 1. Cohort selection and exclusion criteria**

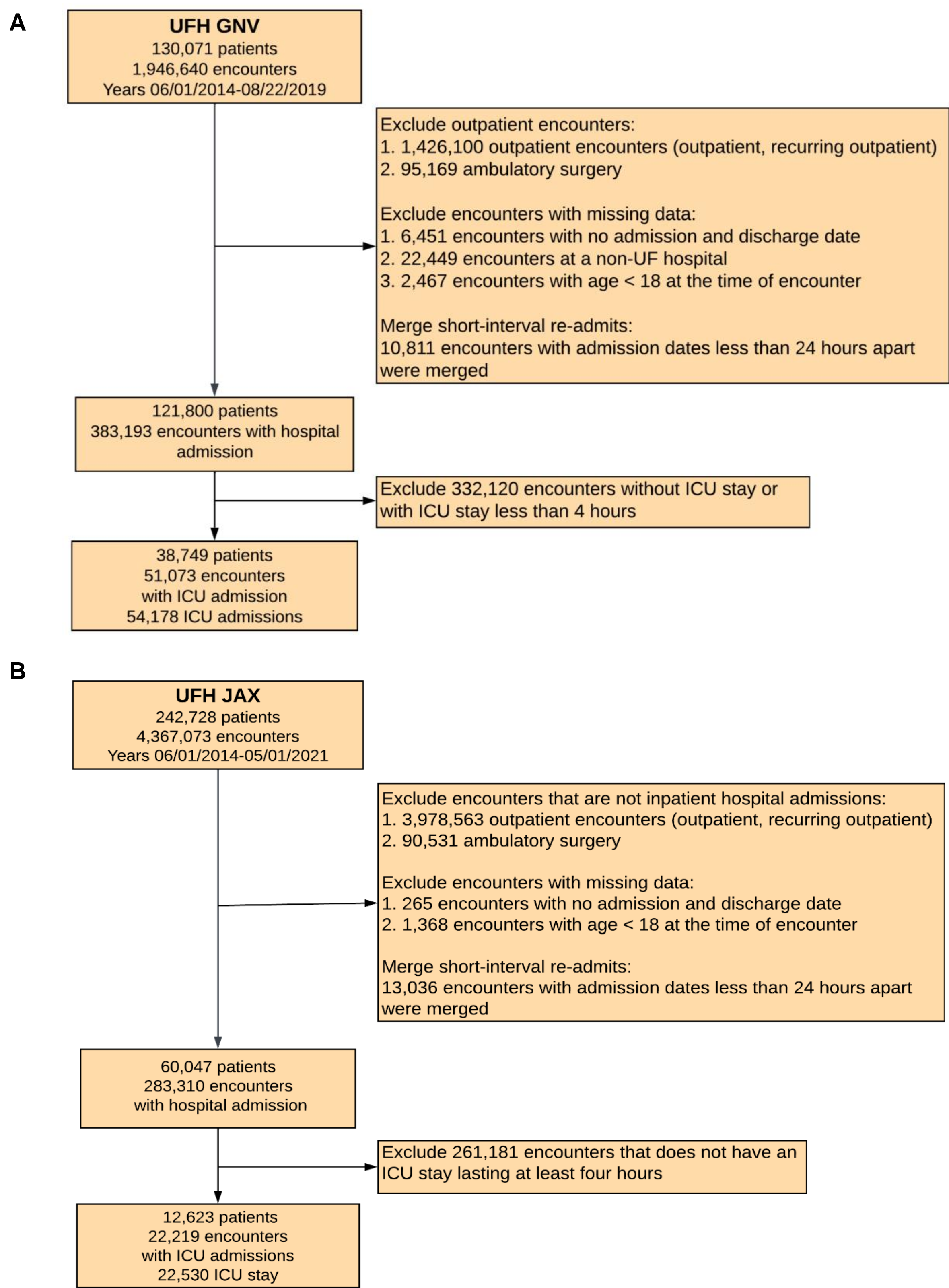

(A) UFH GNV cohort (B) UFH JAX cohort.

**eFigure 2. Flowchart for massive blood transfusion indication**

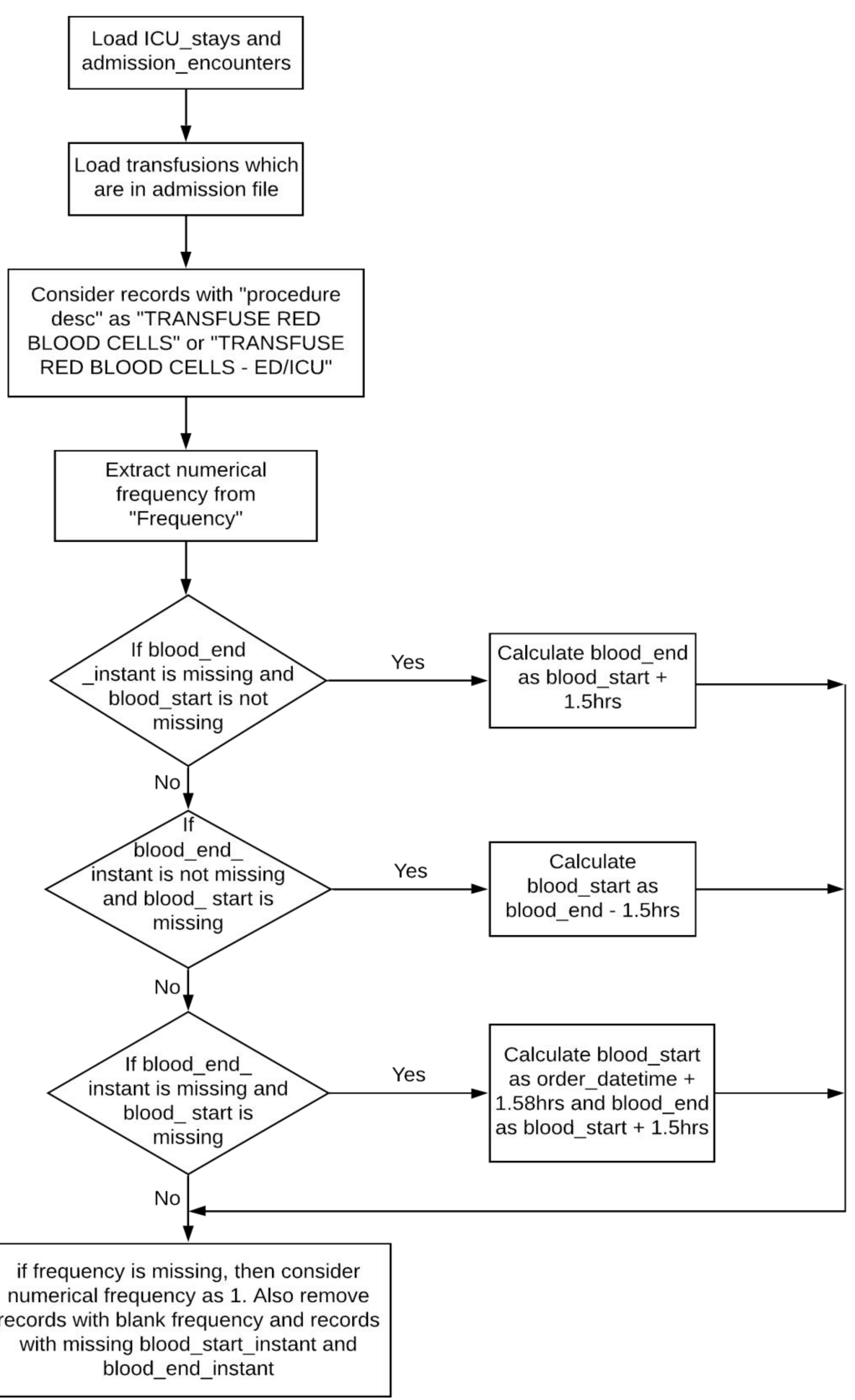

## eFigure 3. Identification of mechanical ventilation use

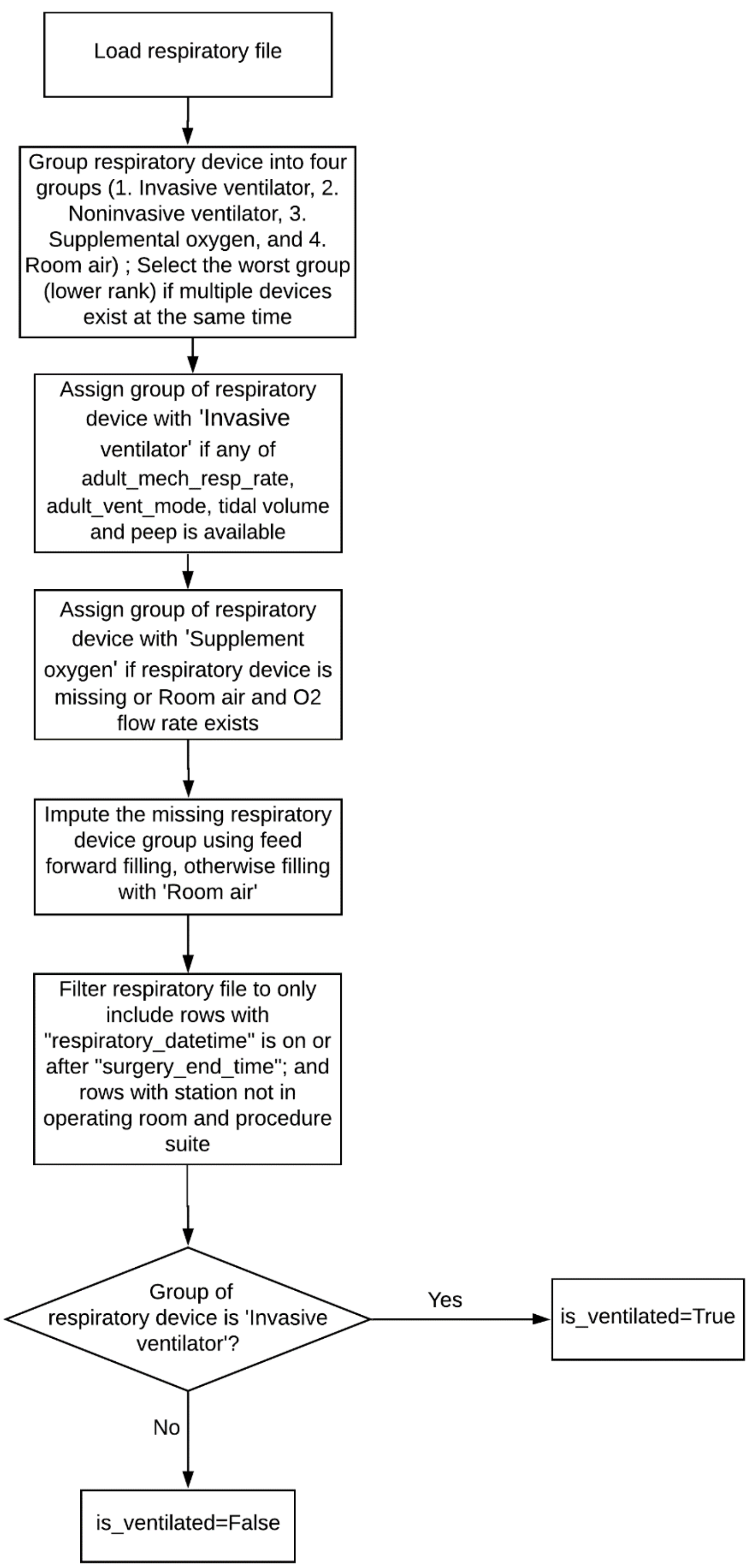

**eFigure 4. Distribution and transition of acuity status for the UFH JAX cohort**

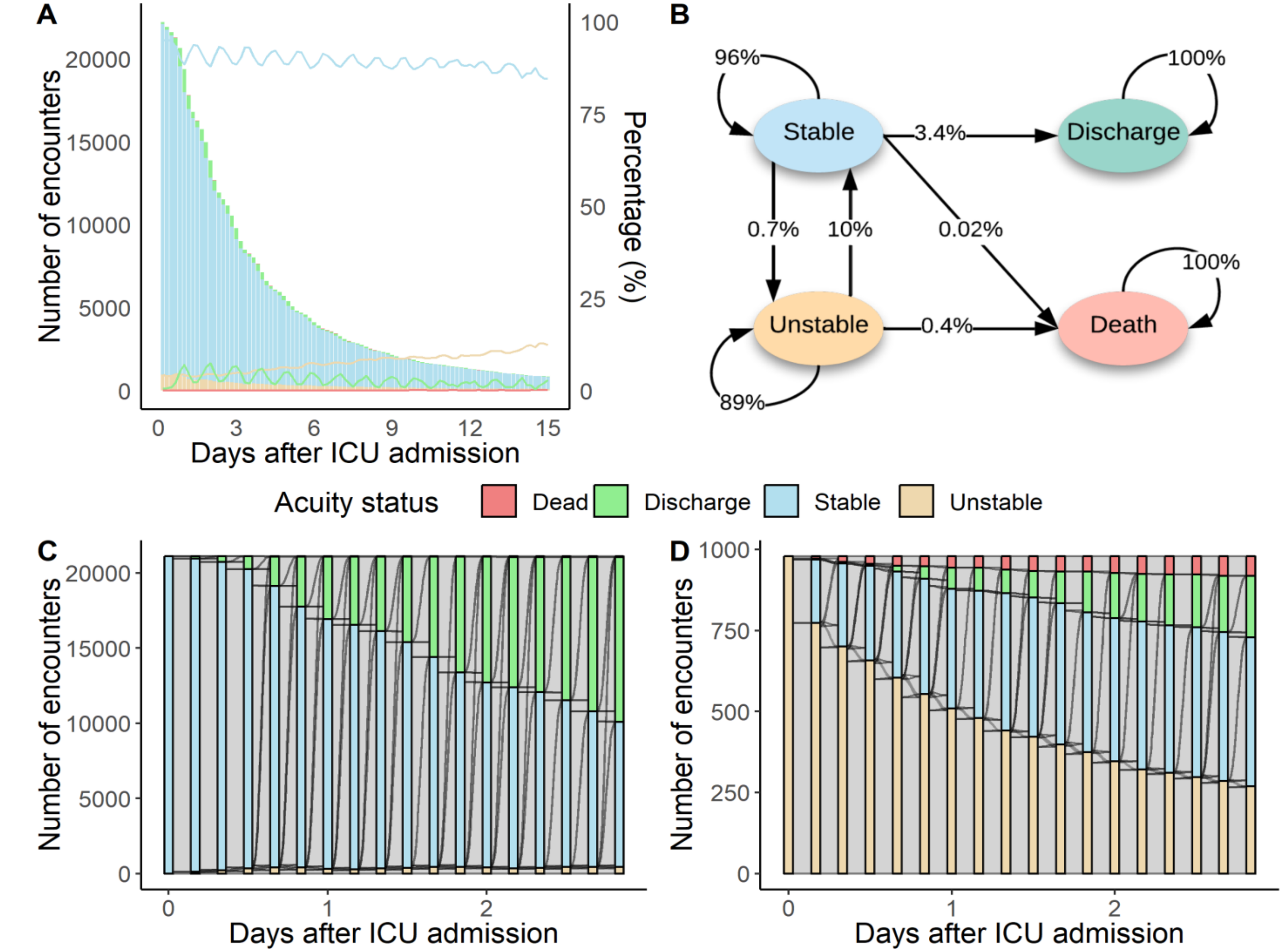

(A) Distribution of acuity status within 15 days of ICU admission. The left y-axis shows the total number of encounters for all acuity states at each time point, which is represented as a bar graph. The right y-axis shows the percentage of each acuity state at each time point, which is represented as a line graph. (B) Acuity state transition probability for every four-hour interval. (C) Acuity status transition for patients with stable status on ICU admission. (D) Acuity status transition for patients with unstable status on ICU admission.

**eFigure 5. Phenotypes of acuity status transition and comparison to Sequential Organ Failure Assessment (SOFA) score for the UFH JAX cohort**

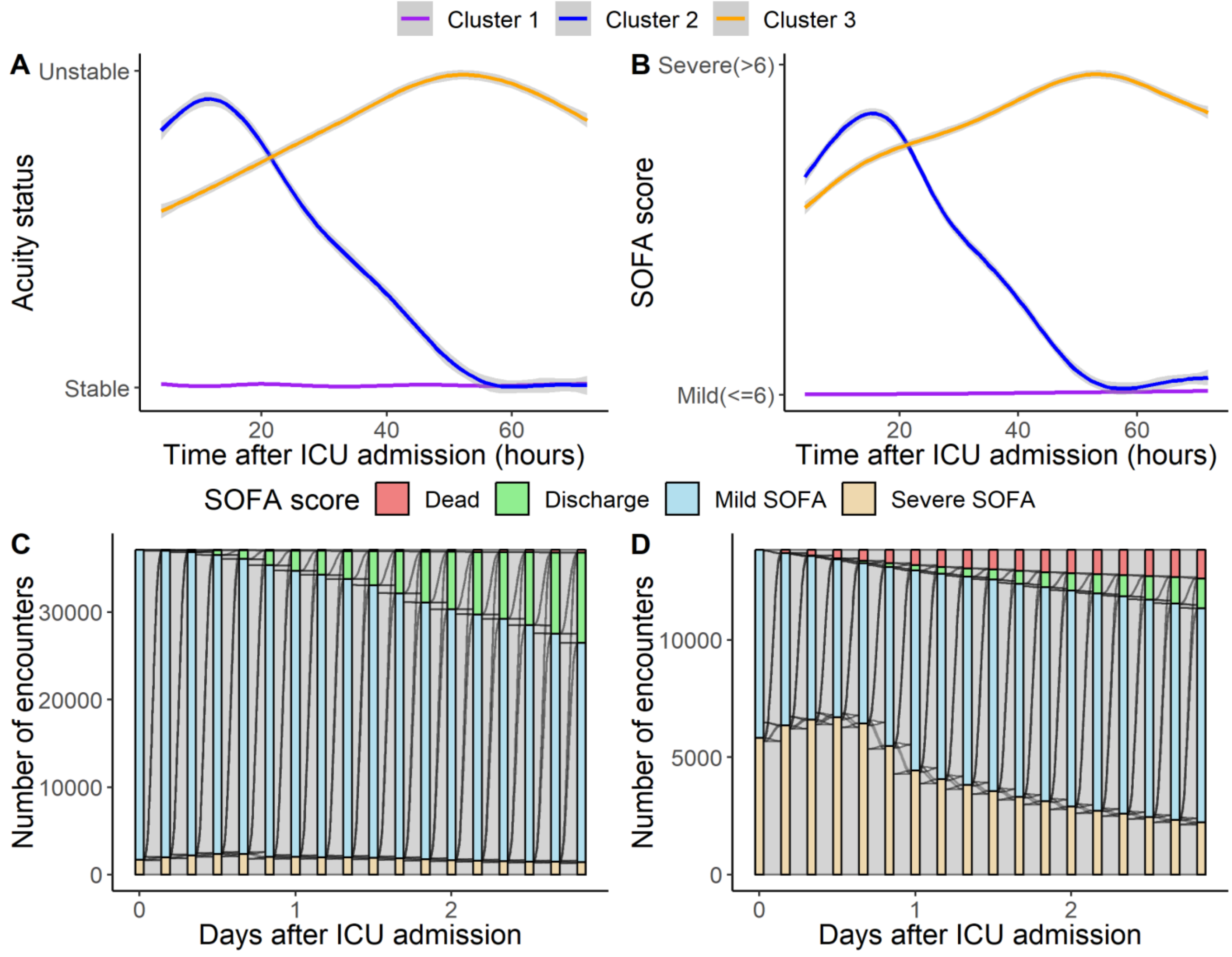

(A) Phenotypes of acuity status transition within the first three days of ICU admission. (B) Phenotypes of SOFA score transition within the first three days of ICU admission. (C) SOFA score transition for patients with stable status on ICU admission. (D) SOFA score transition for patients with unstable status on ICU admission.

**eFigure 6. Distribution of Sequential Organ Failure Assessment (SOFA) across acuity status phenotypes for the UFH GNV cohort**

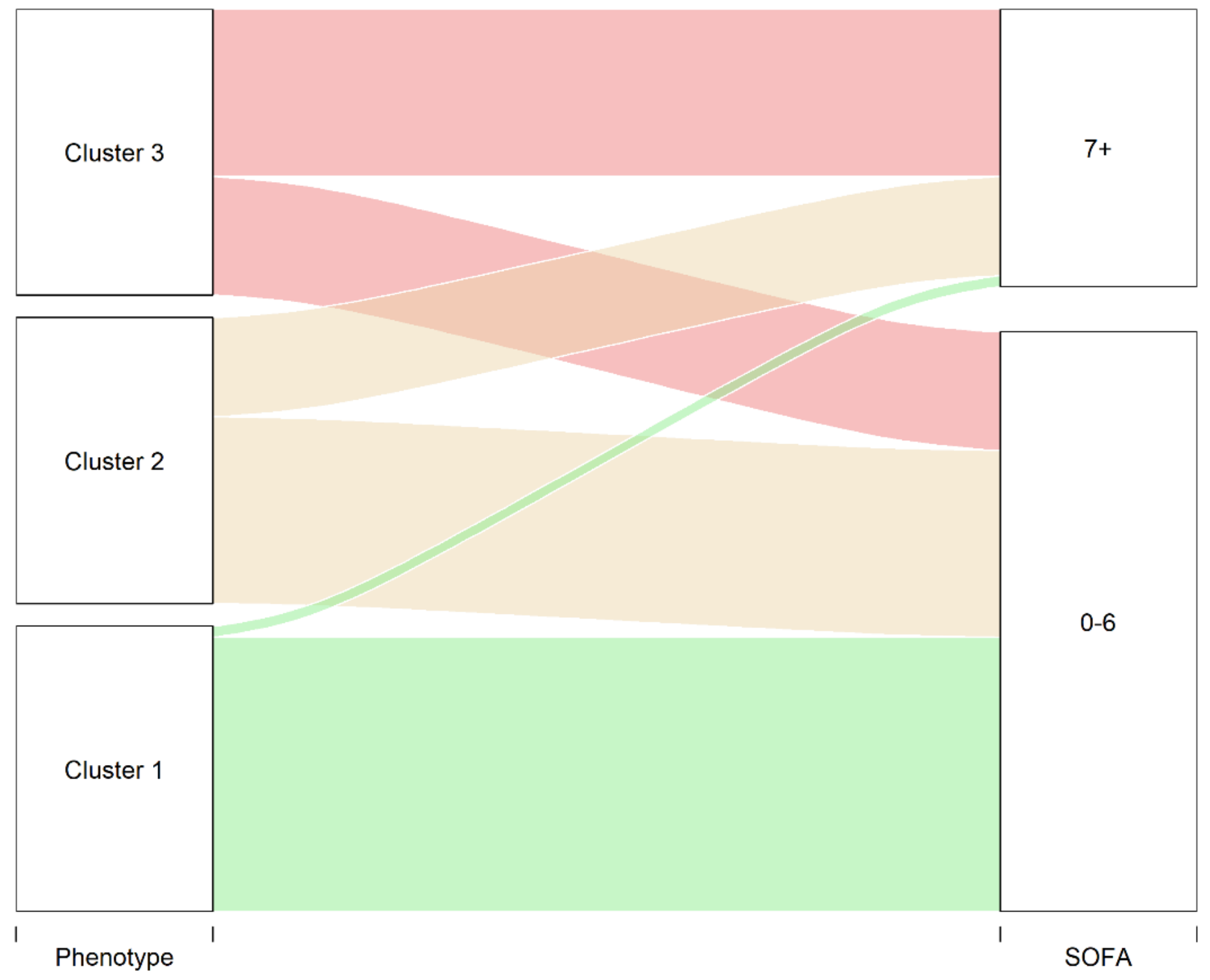

For each phenotype, the larger percentage of patients with that score, the broader the ribbon.

**eFigure 7. Distribution of Sequential Organ Failure Assessment (SOFA) across acuity status phenotypes for the UFH JAX cohort**

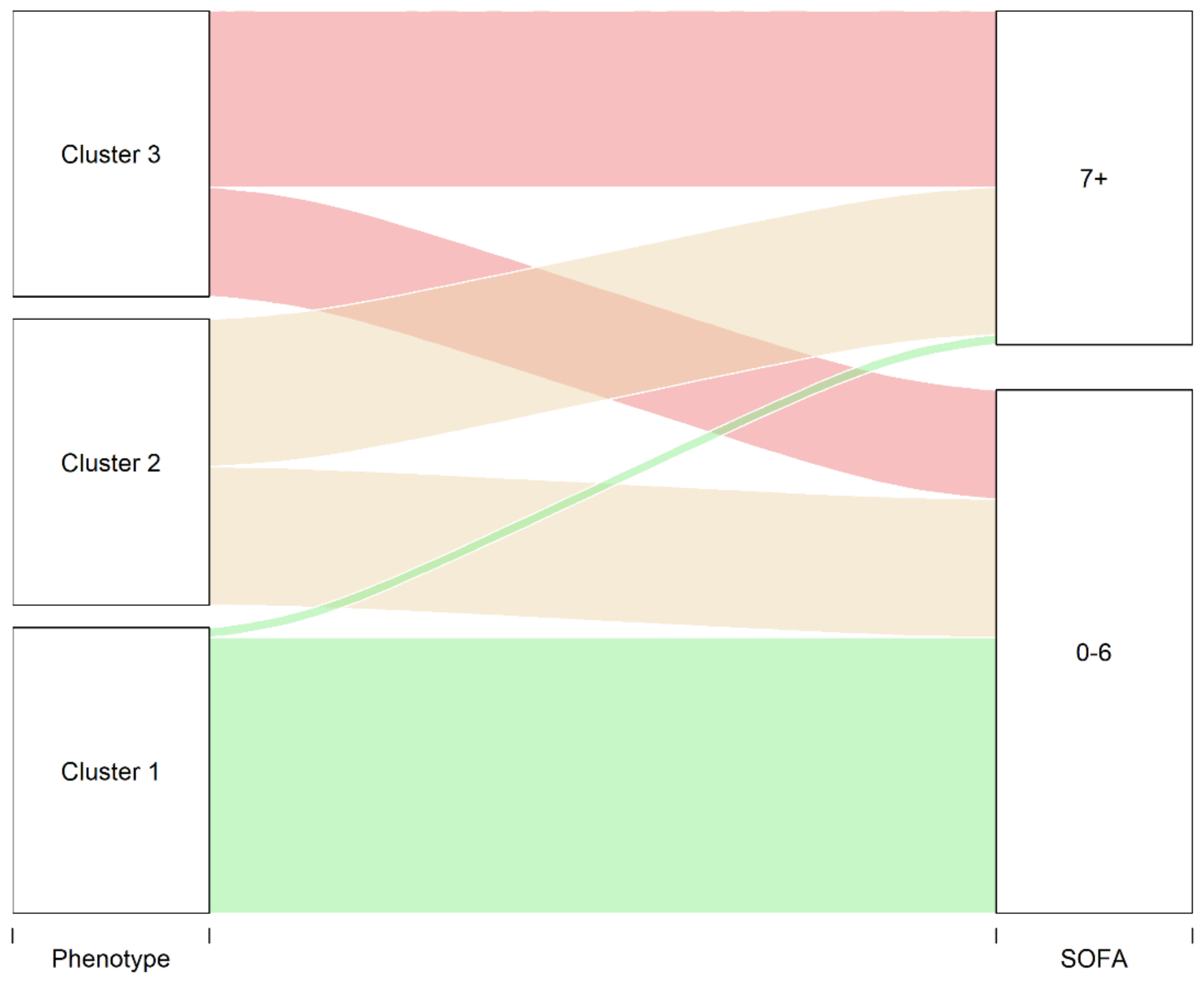

For each phenotype, the larger percentage of patients with that score, the broader the ribbon

**eTable 1. Data elements in the acuity phenotyping algorithm.**

| Used for identification of | Features | Description | Format |
|---|---|---|---|
| Identifier | patient_deiden_id | Deidentified Patient ID | Strings |
| Identifier | encounter_deiden_id | Deidentified Encounter ID | Strings |
| Continuous renal replacement therapy | meas_value | Measured Value | Measured Value |
| Continuous renal replacement therapy | recorded_time | Recorded Time | The date and time value was recorded |
| Continuous renal replacement therapy | vital_sign_group_name | Name of the measured variable | Group name of vital sign. Values include: Device Number, Hourly Net Balance, Maintenance, Output (mL), Prescription, Therapy, or Treatment. |
| Continuous renal replacement therapy | vital_sign_measure_name | Name of the measured variable | Name of the measured variable |
| Mechanical Ventilation | respiratory_datetime | Respiratory DateTime | The date and time when the respiratory device is used |
| Mechanical Ventilation | respiratory_device | Respiratory Device | The device being used to deliver oxygen or move air in and out of the lungs |
| Mechanical Ventilation | adult_vent_mode | Adult Ventilator Mode | The breathing pattern programmed into the mechanical ventilator, which is moving air in and out of the lungs |
| Mechanical Ventilation | adult_mech_resp_rate | Mechanical Respiratory Rate | Breaths per minute for a patient on a mechanical ventilator/breathing machine |
| Mechanical Ventilation | peep | End of Expiratory Pressure | The pressure in the airways at the end of exhalation (mm Hg or cm H20) |
| Mechanical Ventilation | tidal_volume | Tidal Volume (mL) | The volume of air that moves with each breath (mL) |

| Mechanical Ventilation | etco2 | End-tidal Carbon Dioxide Amount | The amount of CO2 in the air that is moving out of the lungs (mm Hg) for patients on mechanical ventilation and without invasive airway, fio2 |
|---|---|---|---|
| Massive blood transfusion | blood_end_instant | Blood transfusion end date and time | Blood transfusion end date and time |
| Massive blood transfusion | blood_start_instant | Blood transfusion start date and time | Blood transfusion start date and time |
| Massive blood transfusion | frequency | Amount of blood transfusion | Amount of blood transfusion as Transfuse X units |
| Massive blood transfusion | procedure_desc | Description of the procedure | Description of the procedure as Transfuse Red Blood Cells/Plasma/Platelets/… |
| Vasopressor Use | taken_datetime | Action Taken DateTime | Time at which MAR action was logged |
| Vasopressor Use | med_order_display_name | Medication Order Display Name | Medication order display name |
| Vasopressor Use | rxnorm_concat | Concatenated Medication Name | The concatenated medical name of the medicine |
| Vasopressor Use | mar_action | Medical Administration Record Action Taken | Medical administration record action taken |
| Vasopressor Use | med_order_discrete_dose | Medication Order Discrete Dose | The dosage at which the medication needs to be administered |
| Vasopressor Use | med_order_route | Medication Order Route | The medium through which the medication is administered |
| Vasopressor Use | med_order_discrete_dose_unit | Medication Order Discrete Dose Unit | The units of the medication dosage |
| Vasopressor Use | height_weight_datetime | Height and Weight Measured DateTime | The date and time that the patient's height and weight are measured. |
| Vasopressor Use | weight_kgs | Weight (kgs) | Patient's weight in kilograms (kg) |

**eTable 2. Vasopressor identification using RxNorms**

| Vasopressor Name | RxNorm |
|---|---|
| Dopamine | 3628, 82010, 1292887, 1743862, 310013, 1743870, 1292740, 310011, 1743878, 1292751, 310012, 1743939, 1114880, 238218, 352586 |
| Epinephrine | 310132, 362, 3992, 310116, 372030, 1660014, 205871 |
| Norepinephrine | 7508, 242969, 7512 |
| Phenylephrine | 8163, 8164, 1242903, 1232651, 373369, 238230 |
| Vasopressin | 11149, 313578, 374283 |

**eTable 3. Grouping of respiratory devices**

| Respiratory device group | Respiratory device |
|---|---|
| Invasive ventilator | ventilator, blow-by, oscillator, tracheostomy, cricothyrotomy, t-piece, bvm (bag valve mask), ett (endotracheal tube), lma (laryngeal mask airways), king tube, transtracheal catheter |
| Noninvasive ventilator | oxyimiser, bipap, oxyhood, cpap, high flow nasal cannula |
| Supplemental oxygen | nasal cannula, trach mask, venturi mask, aerosol mask, non-rebreather mask, simple mask, face tent, partial rebreather mask |
| Room air | room air |

**eTable 4. Clinical outcomes across phenotypes of acuity status and SOFA score for the UFH JAX cohort**

| **Acuity status phenotypes** | **Cluster 1** | **Cluster 2** | **Cluster 3** |
|---|---|---|---|
| Number of encounters, n (%) | 21,095 (95) | 436 (2) | 688 (3) |
| Length of hospital stay, days, median (IQR) | 4 (2, 7) | 7 (4, 11)[a] | 11 (5, 21)[a,b] |
| Length of ICU stay, days, median (IQR) | 3 (2, 5) | 4 (2, 8)[a] | 6 (2, 13)[a] |
| Hospital mortality, n (%) | 313 (1) | 13 (3) | 238 (36)[a,b] |
| Death within 7 days of ICU admission, n (%) | 110 (0.5) | 5 (1) | 167 (24)[a,b] |
| **SOFA score phenotypes** | **Cluster 1** | **Cluster 2** | **Cluster 3** |
| Number of encounters, n (%) | 13,707 (62) | 6,667 (30) | 1,845 (8) |
| Length of hospital stay, days, median (IQR) | 4 (2, 6) | 5 (3, 10)[a] | 8 (4, 15)[a] |
| Length of ICU stay, days, median (IQR) | 3 (2, 4) | 4 (2, 6)[a] | 4 (2, 9)[a,b] |
| Hospital mortality, n (%) | 86 (1) | 170 (3)[a] | 308 (17)[a,b] |
| Death within 7 days of ICU admission, n (%) | 38 (0.3) | 61 (0.9)[a] | 183 (10)[a,b] |

Abbreviations: SOFA: sequential organ failure assessment; ICU: intensive care unit; IQR: interquartile range.
All p-values were adjusted for multiple comparisons using Bonferroni method.
[a] $p < 0.05$ compared to Cluster 1. .
[b] $p < 0.05$ compared to Cluster 2.

**eTable 5. Clinical outcomes of patients with mild sequential organ failure assessment score (<6) across phenotypes of acuity status on UFH GNV and JAX cohorts.**

| **UFH GNV cohort** | | | |
|---|---|---|---|
| **Acuity status phenotypes** | **Cluster 1 (N=37,180)** | **Cluster 2 (N=6,219)** | **Cluster 3 (N=7,674)** |
| Number of encounters with mild SOFA score, n (%) | 35,730 (96) | 4,064 (65) | 3,178 (41) |
| Length of hospital stay, days, median (IQR) | 6 (4, 10) | 8 (5, 12)[a] | 13 (8, 21)[a,b] |
| Length of ICU stay, days, median (IQR) | 3 (2, 5) | 4 (3, 6)[a] | 9 (5, 16)[a,b] |
| Hospital mortality, n (%) | 1,232 (3) | 133 (3) | 906 (29)[a,b] |
| Death within 7 days of ICU admission, n (%) | 517 (1) | 58 (1) | 590 (19)[a,b] |
| **UFH JAX cohort** | | | |
| **SOFA score phenotypes** | **Cluster 1 (N=21,095)** | **Cluster 2 (N=436)** | **Cluster 3 (N=688)** |
| Number of encounters with mild SOFA score, n (%) | 20,369 (97) | 211 (48) | 263 (38) |
| Length of hospital stay, days, median (IQR) | 4 (2, 7) | 5 (3, 10)[a] | 10 (5, 19)[a,b] |
| Length of ICU stay, days, median (IQR) | 3 (2, 5) | 4 (2, 6)[a] | 4 (2, 10)[a] |
| Hospital mortality, n (%) | 255 (1) | 3 (1) | 60 (23)[a,b] |
| Death within 7 days of ICU admission, n (%) | 85 (0.4) | 0 (0) | 45 (17)[a,b] |

Abbreviations: SOFA: sequential organ failure assessment; ICU: intensive care unit; IQR: interquartile range.
All p-values were adjusted for multiple comparisons using Bonferroni method.
[a] $p < 0.05$ compared to Cluster 1. .
[b] $p < 0.05$ compared to Cluster 2.